\documentclass{epl}
\newcommand{\be}{\begin{equation}}
\newcommand{\ee}{\end{equation}}

\title{Sufficient Conditions for Topological Order in Insulators}
\author{M. B. Hastings}
\institute{
Center for Nonlinear Studies and Theoretical Division, Los Alamos National
Laboratory, Los Alamos, NM 87545, hastings@lanl.gov 
}
\date{November 1, 2004}
\begin{document}
\maketitle
\begin{abstract}
We prove the existence of low energy excitations in insulating systems at
general filling factor under certain conditions, and discuss in which
cases these may be identified as topological excitations.  
In the specific case of half-filling this proof
provides a significantly shortened proof of the recent
higher dimensional Lieb-Schultz-Mattis theorem.
\vskip2mm
\end{abstract}
The classic 1961 result of Lieb, Schultz, and Mattis (LSM)\cite{lsm}, proving
the existence of an excitation within energy $\sim 1/L$ of the ground state
for certain one dimensional spin chains, has had a large effect on the field.
While it was then proven by Affleck and Lieb\cite{td} that one dimensional
systems either have gapless localized excitations or a local symmetry
breaking, it has long been suspected that in higher dimensions there is a
more interesting possibility of topological order\cite{topo}.

One way to understand topological order is based on flux insertion.
We give the physical argument here, and then discuss the difficulties
in this argument which give rise to the need for the more careful
argument of this paper.
We consider a higher-dimensional system which is periodic in one direction.
A spin-$1/2$ system can be mapped to a hard-core boson system on a lattice, with
the presence or absence of a particle denoting spin up or down.
If the particle system is superfluid, there is long range order in the
$x$ and $y$ components of the spin in the original system, implying the
existence of low energy excitations.  On the other hand, if the particle
system is insulating,
it should be possible
to insert $2\pi$ of gauge flux in the hopping of particles
across a given line cutting the system, returning the Hamiltonian to
the original one, but, for non-integer filling fraction,
taking the system to an excited state which is
very close in energy to the ground state.  Using adiabatic flux
insertion, this was suggested as a way to prove a higher dimensional LSM
theorem\cite{oshi}.  The two possibilities would thus seem to be a superfluid
system (or other system which resists flux insertion) which has low
energy excitations, or an insulating system which has topological order.
In either case, there is a state close in energy to the ground state.

However, there is a serious problem with this argument.
The definition of adiabatic flux insertion depends on the existence
of a gap (but does not require anything about the magnitude of
a gap); however, in a spin system
with no disorder,
any gap in the
spectrum at zero flux must close at some non-zero value of the flux, thus
making it impossible even to define an adiabatic
flux insertion\cite{ml}.  
In a fractional quantum Hall system with disorder, 
there is a related problem.  In a $1/3$ quantum Hall state,
the gap between the three approximately degenerate ground states
remains open for all values of the flux, even though the gap is
exponentially small.  However, this means that adiabatic flux insertion
leaves the system in the lowest of the three states and does not produce
the correct topologically excited state\cite{fqhe}.
Thus flux must be inserted slowly, but not adiabatically.  For fractional
Hall systems, the experimentally relevant flux insertion is slow enough
to avoid exciting states above the three almost degenerate ground states,
but fast enough to ``shoot through" the level crossing.

A precise definition of this quasi-adiabatic flux insertion
was given in \cite{lsmh}, and
used to prove a version
of the LSM theorem for spin systems with total $S^z=0$
valid in arbitrary dimension.  The flux was inserted in
such a way that the ground state wave function was only disturbed near the
flux insertion point.  Physically, one can imagine that the flux was
inserted sufficiently rapidly to prevent influences propagating around the
system, but sufficiently slowly to avoid creating localized excitations.
Compared to \cite{lsmh}, the proof here
is shorter and
is generalized to
systems at general filling fraction $\rho$ instead
of just spin systems with total $S^z=0$.  It also
explicitly constructs the
low energy states for certain systems.
Later, we discuss under what conditions
these states can be identified as topological excitations.

{\it Definition of System---}
We consider systems defined on a lattice, using letters $i,j,...$
to denote lattice sites (throughout, the term ``site" may also be
used to refer to a unit cell comprised of
several sites), and introducing a metric $d(i,j)$ on the lattice.
We assume that there is a conserved charge,
$Q=\sum_i Q_i$, where $Q_i$, the charge on site $i$, is quantized to be an
integer (for example, $Q_i$ may be taken to be the number of particles on a
site, or in a spin system of half-integer spins
$Q_i$ may be equal to $S^z_i+1/2$).  We assume that the Hamiltonian, ${\cal H}$ can be written
as ${\cal H}=\sum_i {\cal H}^i$, where the ${\cal H}^i$ obey the finite
range conditions\cite{fgv,loc}: $(1)$ the commutator $[{\cal H}^i,O]=0$ for
any operator $O$ which acts only on sites $j$ with $d(i,j)>R$ where $R$
is the range of the Hamiltonian; and $(2)$ the operator norm $||{\cal H}^i||
\leq J$ for all $i$, for some constant $J$.  Finally, the
number of sites $j$ with $d(i,j)\leq R$ should be bounded by some number that
we denote
$S$.  These finite range
conditions include all short-range
boson systems with a {\it finite} number of bosons allowed per site, as
well as all short-range spin systems.

We suppose that there are a total of $V$ sites on the $d$-dimensional
lattice.  We assume that the Hamiltonian is translationally invariant
in one of the directions of the lattice, with period $L$: this is the
length of the system.
We define the filling fraction of the system
to be $\rho=Q/V$.
For each site $i$, we introduce coordinates $(x_i,y_i)$ to specify sites, where
$x_i$
labels the distance along the length direction and $y_i$ labels the transverse
directions.  The coordinate $x$ is periodic with period $L$ and should
be compatible with the metric $d(i,j)$:
any two sites $i,j$ should have $d(i,j)$ greater
than or equal to the minimum over integers $n$ of $|x_i-x_j-nL|$.

We also define a twisted Hamiltonian and a rotation
operator.  Let 
\be
R(\theta)=\prod_{j,0<x_j \leq L/2}
\exp[i\theta (Q_j-\rho)],
\ee
where the product
ranges over all sites $j$ such that $0<x_j\leq L/2$. 
We introduce
two boundary condition twists as follows.  Let 
${\cal H}_{\theta_1,\theta_2}=\sum_i {\cal H}^i_{\theta_1,\theta_2}$, where
${\cal H}^i_{\theta_1,\theta_2}=R(\theta_1){\cal H}^i R(-\theta_1)$ if
$-R\leq x_i\leq R$ and
${\cal H}^i_{\theta_1,\theta_2}=R(-\theta_2){\cal H}^i R(\theta_2)$ if
$L/2-R \leq x_i \leq L/2+R$ and otherwise
${\cal H}^i_{\theta_1,\theta_2}={\cal H}^i$.  Thus,
angle $\theta_1$ defines a twist in boundary conditions between $x=0$
and $x=1$, while
angle $\theta_2$ defines a twist between $x=L/2$ and $x=L/2+1$.
These twists correspond to inserting gauge flux between $x=0$ and
$x=1$ or between $x=L/2$ and $x=L/2+1$.
Let there be a total of
$N_c$ sites $i$ with $-R\leq x_i \leq R$, and let
$||\partial_\theta{\cal H}^i_{\theta_1,\theta_2}||\leq K$ for all $i$ for
some constant $K$.
For a $d$ dimensional
system with linear size $L$ and aspect ratio close to unity, $N_c\sim L^{d-1}$.

Let $\Psi_0(\theta_1,\theta_2)$ be the ground state of ${\cal H}_{\theta_1,\theta_2}$,
and let $\Psi_0$ be the ground state of ${\cal H}_{0,0}={\cal H}$.
Note that ${\cal H}_{\theta_1,\theta_2}=R(-\theta_2)
{\cal H}_{\theta_1+\theta_2,0}R(\theta_2)$ and in particular
${\cal H}_{\theta,-\theta}=R(\theta){\cal H}R(-\theta)$
and $\Psi_0(\theta,-\theta)=R(\theta)\Psi_0$.
The second boundary condition angle thus appears to be redundant, since
it is only the sum of the two angles, $\theta_1+\theta_2$, that
affects the eigenspectrum of ${\cal H}_{\theta_1,\theta_2}$.  That is,
the physical properties depend only on the total gauge flux
$\theta_1+\theta_2$.  However,
the introduction of the second boundary angle will be extremely useful below.

{\it Results---}
The main result is that for a short-range Hamiltonian as defined
above we are able to construct topologically excited states
under certain assumptions on the spectrum of the Hamiltonian.
The topologically excited states have a low energy but momentum that
differs from the ground state momentum as given in
Eq.~(\ref{res1},\ref{res2},\ref{eno}).
We begin by specifying the conditions on the spectrum.

Suppose that, for $\theta_1=\theta_2=0$, there is a gap
$\Delta E$ in energy between the ground state and a set of eigenstates,
that we refer to as local excitations; we define
$P_{high}$ to be the projection operator onto this space of states.
There may be other eigenstates
with energy below this
gap; we 
define
$P_{low}=1-P_{high}$ to project onto this space.  We
assume that for certain local operators $O$,
namely $O=\partial_{\theta}{\cal H}_{\theta,-\theta}$ and
$O=\partial_\theta R(\theta)$,
the matrix elements of these operators, between $\Psi_0$ and any other
normalized state $\Psi$ which is a linear combination of eigenstates
below the gap and which is orthogonal to the ground state,
are bounded by $\epsilon||O||$ for some small constant $\epsilon$.
Physically, we are interested in this case because in many situations
the matrix elements of local operators
between the ground state and topologically excited
states are very small, often exponentially small in the system size.
Since the Hamiltonian is translationally invariant, the ground state, $\Psi_0$,
may be taken translationally invariant, with eigenvalue $z_0$:
$T\Psi_0=z_0\Psi_0$ where $T$ is the translation operator which
increases the $x$-coordinate by 1.

Then, we introduce the twist operator
\begin{eqnarray}
\label{weq}
W_1(\phi)=\Theta\exp\{-\int_0^{\phi}
{\rm d}{\theta}\int_0^{\infty} {\rm d}\tau 
\exp[-(\tau\Delta E)^2/(2q)]
[\tilde w_{1,\theta} ^+(i\tau)-
h.c.]
\},
\end{eqnarray}
where the symbol $\Theta$ denotes that the exponential is $\theta$-ordered,
in analogy to the usual time ordered or path ordered exponentials.  We
define $w_{1,\theta}=\partial_\theta{\cal H}_{\theta,0}$ and define
$\tilde w_{1,\theta}^+(i\tau)$
following \cite{loc}: for
any operator $A$
\begin{eqnarray} 
\label{tdef}
{\tilde A}(t)\equiv A(t) 
\exp[-(t\Delta E)^2/(2q)]
; \;
{\tilde A}^{\pm}(\pm i\tau)
=\frac{1}{2\pi}
\int {\rm d}t \,
{\tilde A}(t)\frac{1}{\pm it+\tau}.
\end{eqnarray}
The time evolution of operators is defined by
$A(t)=\exp[i{\cal H}_{1;\theta}t]A\exp[-i{\cal H}_{1;\theta}t]$, where
we define the Hamiltonian ${\cal H}_{1;\theta}=\sum_i {\cal H}^i_{\theta,0}$,
with the sum ranging over $-L/4+R<x_i<L/4-R$.  
The Hermitian conjugate in Eq.~(\ref{weq}) of
$\tilde w_{1,\theta}^+(i\tau)$ 
is
$\tilde w_{1,\theta}^-(-i\tau)$, and $W_1(\theta)$ is a unitary operator.
We define $W_1=W_1(2\pi)$.

To understand the definition (\ref{tdef}),
for any operator $A$ define
$A^+(i\tau)$ to be the positive energy part of $A$
at imaginary time $i\tau$.  That is, in a basis of eigenstates of
the Hamiltonian with energies $E_i$, we have
$A^+_{ij}(i\tau)=A_{ij} \exp[-\tau (E_i-E_j)] 
\Theta(E_i-E_j)$, with the step function $\Theta(x)=1$ for $x>0$, 
$\Theta(0)=1/2$, and $\Theta(x)=0$ for $x<0$.  The operator
$\exp[-(t\Delta E)^2/(2q)]\tilde A^{+}(i\tau)$ is a good approximation
to $A^+(i\tau)$ in the following sense:
the difference
$|P_{high}\exp[-(\tau\Delta E)^2/(2q)]\tilde A^{+}(i\tau)\Psi_0-
P_{high} A^+(i\tau)\Psi_0|$ is bounded by $\exp[-(\tau \Delta E)^2/(2q)]
\exp[-q/2] ||A||$ for $\tau<q/\Delta E$ and is bounded by
$\exp[-\tau \Delta E]||A||$ for $\tau>q/\Delta E$.  This result follows
from elementary integrations\cite{loc} and is a kind of
uncertainty relation: $\tilde A(t)$ is cutoff by the Gaussian at
times of order $1/\Delta E$ and so can be used to approximate states
with energies of order $\Delta E$.
In the limit $q\rightarrow\infty$, when the approximation $\tilde A^+(i\tau)$
becomes exact, Eq.~(\ref{weq}) adiabatically twists $\theta_1$ by $\phi$;
we instead keep $q$ finite as chosen later.

The use of
${\cal H}_{1;\theta}$ to define the time evolution, rather
than ${\cal H}_{\theta,0}$ as in \cite{lsmh} is a very useful
technical trick which simplifies the proof.  Using ${\cal H}_{1;\theta}$,
we have $[W_1,O]=0$ for any operator $O$ which acts only on a
sites $j$ with $L/4\leq x_j\leq 3L/4$.  That is, $W_1$
acts only on sites $j$ with
$-L/4<x_j<L/4$.

We define
$\Psi_n=W_1^n\Psi_0$ and define $E_n=
\langle \Psi_n | {\cal H} | \Psi_n \rangle-\langle {\cal H} \rangle$
where 
the first and second pairs of
angle brackets denote the inner product and ground state expectation value
respectively.
We prove below that
\be
\label{res1}
E_n \leq 
VJ e_n/2,
\ee
where
$e_n$ is given by Eq.~(\ref{end})
and 
\be
\label{res2}
|\langle \Psi_n|T|\Psi_n\rangle-\langle T \rangle 
\exp[i2 \pi n\rho(V/L)]|\leq e_n.
\ee

For large $L$, we will find that $e_n$ is of order
\be
\label{eno}
e_n \sim (K/\Delta E)[N_c^2 \exp(-c_1 L \Delta E/8)
+N_c^2 \sqrt{c_1 L \Delta E}\exp(-L/4\xi_C)
+N_c \epsilon \sqrt{c_1 L \Delta E}],
\ee
where the constants
$c_1,\xi_C$ are defined below.  Thus, for a $d$-dimensional system
with $N_c\propto L^{d-1}$ and with fixed $\Delta E$ and vanishingly
small $\epsilon$, we have $e_n,E_n$ going to zero exponentially in $L$.

One consequence of these results is a higher dimensional
Lieb-Schultz-Mattis theorem for systems of with $\rho(V/L)$ non-integer,
a case which includes spin systems with half-integer
spin per unit cell, total $S^z=0$, and odd width $V/L$\cite{lsmh}.
If we set $\Delta E$ to be equal to
the gap between the ground state and the first excited state, then $\epsilon=0$
as the ground state is the only state with energy less than $\Delta E$.
Then, if $\Delta E=k\log(L)/c_1 L$,
the exponential $\exp(-c_1 L \Delta E/8)$ decays as $L^{-k/8}$, and
for large enough $k$ the state $\Psi_1$ will be less than $\Delta E$ in
energy, while we can use the expectation
value of $T$ for state $\Psi_1$ to bound the overlap of $\Psi_1$ with
the ground state, leading to a contradiction and thus bounding $\Delta E$.

To show these results, we will need a few additional definitions defined
in the next two sections.
For use later, 
we introduce the operator 
\be
W_2(\phi)=
\Theta\exp\{-\int_0^{-\phi}
{\rm d}{\theta}\int_0^{\infty} 
{\rm d}\tau 
\exp[-(\tau\Delta E)^2/(2q)]
[\tilde w_{2,\theta}^+(i\tau)-h.c.]
\},
\ee
where $w_{2,\theta}=\partial_\theta{\cal H}_{0,\theta}$.
We write $W_2=W_2(2\pi)$.
Here, the time evolution of operators is
defined using the Hamiltonian ${\cal H}_{2;\theta}\equiv
\sum_i {\cal H}^i_{0,\theta}$,
with the sum ranging over $L/4+R<x_i<3L/4-R$.
Finally, we define
$W(\phi)=\Theta\exp\{-\int_0^{\phi}
{\rm d}{\theta}\int_0^{\infty}
{\rm d}\tau 
\exp[-(\tau\Delta E)^2/(2q)]
[\tilde w_{\theta}^+(i\tau)-
h.c.]
\}$, where $w_{\theta}=\partial_{\theta}{\cal H}_{\theta,-\theta}$ and
where the time evolution is defined using the Hamiltonian
${\cal H}_{\theta,-\theta}$.

{\it Twist in Boundary Condition---}
This and the next section show two basic facts about $W,W_1,W_2$.
First,
the difference $|W(\phi)\Psi_0-R(\phi)\Psi_0|$ is exponentially small
in $q$.  Thus,
$W$ ``twists"
the boundary conditions, twisting $\theta_1$ by $\phi$ and $\theta_2$
by $-\phi$.  Second, the
difference $||W_1(\phi)W_2(\phi)-W(\phi)||$ is exponentially small
in $(c_1 L\Delta E)^2/q$,
and thus one can approximately factor the
operator $W$ into the product of two commuting operators $W_1,W_2$.
Physically, the operator $W_1(\phi)$ twists $\theta_1$ by $\phi$ while
$W_2(\phi)$ twists $\theta_2$ by $-\phi$.
We will then be able to pick a $q$ of order $L$ such that both differences are
exponentially small in $L$.  Given these two results, 
Eqs.~(\ref{teq},\ref{cbeq}), it will be easy to
derive the main results (\ref{res1},\ref{res2})
above.  The reader who would prefer to
skip the derivation of these two results may skip to the section
``Bound on Energy".

We start by using linear perturbation theory to compute
$\partial_{\theta}\Psi_0(\theta,-\theta)=
\partial_{\theta}R(\theta)\Psi_0$.
We have 
${\cal H}_{\theta,-\theta}R(\theta)\Psi_0=E_0 R(\theta)\Psi_0$.
Taking derivatives of both sides with respect to $\theta$ at $\theta=0$,
and working in a basis of eigenstates $\Psi_i$ of ${\cal H}$ with energies
$E_i$, we can compute the matrix element
$(\partial_{\theta} R(\theta))_{i0}$ if
$E_i\neq E_0$:
$(\partial_{\theta} R(\theta))_{i0}=
-(E_i-E_0)^{-1} (\partial_{\theta}{\cal H}_{\theta,-\theta})_{i0}$.
Thus, 
\begin{eqnarray}
\label{lth}
\partial_{\theta}\Psi_0(\theta,-\theta)=
-\sum_{i,E_i\neq E_0}
\frac
{(\partial_{\theta}{\cal H}_{\theta,-\theta})_{i0}}
{E_i-E_0}
\Psi_i
+Z_{i0}\Psi_i
=
-\int_0^{\infty}{\rm d}\tau
[(\partial_{\theta}
{\cal H}_{\theta,-\theta})^+(i\tau)-h.c.]\Psi_0+
Z\Psi_0,
\end{eqnarray}
where we define the operator $Z$ by $Z_{ij}=(\partial_{\theta}R(\theta))_{ij}$
if $E_i=E_j$ and $Z_{ij}=0$ otherwise.
The second
equality in (\ref{lth})
may be verified by computing the integral over $\tau$.
In the limit $q\rightarrow\infty$, the operator
$-\int_0^{\infty}{\rm d}\tau
[(\partial_{\theta'}
{\cal H}_{\theta',-\theta'})^+(i\tau)-h.c.]$
equals
the integrand in the exponent of $W$ and so in that limit,
for a non-degenerate ground
state, 
$W(\phi)\Psi_0=R(\phi)\Psi_0$.

We now consider the case of finite $q$ and
bound the difference $|W(\phi)\Psi_0-R(\phi)\Psi_0|$.
We use
$\partial_{\theta}
{\cal H}_{\theta,-\theta}=R(\theta)
\partial_{\theta'}
{\cal H}_{\theta',-\theta'}R(-\theta)$, taking the derivatives
at $\theta'=0$,
to rewrite
$W(\phi)=\Theta\exp\{-\int_0^{\phi}
{\rm d}{\theta}\int_0^{\infty} 
{\rm d}\tau 
\exp[-(\tau\Delta E)^2/(2q)]
R(\theta)[\tilde w_{0}^+(i\tau)-h.c.]
R(-\theta)
\}$
, where the
time evolution of $w_0$
is defined using the Hamiltonian ${\cal H}_{0,0}={\cal H}$.
Thus, $W(\phi)
=R(\phi)\exp\{-
\phi
\int_0^{\infty} 
{\rm d}\tau 
\exp[-(\tau\Delta E)^2/(2q)]
[\tilde w_{0}^+(i\tau)-
h.c.]-
\phi\partial_{\theta} R(\theta)
\}$, taking the derivatives at $\theta=0$.

Then, $|W(\phi)\Psi_0-R(\Phi)\Psi_0|=
|R(-\phi)W(\phi)\Psi_0-\Psi_0|$ and
\be
R(-\phi)W(\phi)\Psi_0=
\exp{\bf (}-\phi\int_0^{\infty}{\rm d}\tau
\{\exp[-(\tau\Delta E)^2/(2q)]\tilde w_0^+(i\tau)-w_0^+(i\tau)-h.c.\}-\phi Z
{\bf )}\Psi_0.
\ee
We now bound the norm of the state
$\Psi\equiv (-\int_0^{\infty} {\rm d}\tau 
\{ \exp[-(\tau\Delta E)^2/(2q)]\tilde w_0^+(i\tau)-w_0^+(i\tau)-h.c.
\}-Z)\Psi_0$.  
The norm $|\Psi|\leq |P_{high}\Psi|+|P_{low}\Psi|$.  We have
$P_{high}Z\Psi_0=0$, while using the bounds below Eq.~(\ref{tdef})
$|P_{high}(-\int_0^{\infty} {\rm d}\tau 
\{ \exp[-(\tau\Delta E)^2/(2q)]\tilde w_0^+(i\tau)-w_0^+(i\tau)-h.c.
\})\Psi_0|\leq
N_c (K/\Delta E)(2\exp[-q]+\exp[-q/2]\sqrt{2\pi q})$.  

On the other hand, $P_{low}\Psi=
P_{low}[-\int_0^{\infty} {\rm d}\tau 
\exp[-(\tau\Delta E)^2/(2q)]\tilde w_0^+(i\tau)-h.c.-\partial_{\theta}
R(\theta)]\Psi_0$.  The expectation values $\langle \partial_\theta
R(\theta) \rangle$
and $\langle \partial_{\theta} {\cal H}_{\theta,-\theta} \rangle$ both
vanish, due to the translation symmetry of the ground state, so
we may assume that $P_{low} \Psi$ is orthogonal to $\Psi_0$.
Then, we can use the bound on the matrix elements between $\Psi_0$ and
other states below the gap to show that
$|P_{low}\Psi| \leq \epsilon \sqrt{2\pi q} N_c (K/\Delta E)
+\epsilon ||\partial_{\theta'} R(\theta')||$.  
Therefore, 
$|\Psi|\leq
N_c (K/\Delta E)[2\exp(-q)
+\exp(-q/2)\sqrt{2\pi q}]+
\epsilon \sqrt{2\pi q} N_c (K/\Delta E)
+\epsilon ||\partial_{\theta'} R(\theta')||$.

Therefore,
$|W(\phi)\Psi_0-R(\phi)\Psi_0|\leq c_2(\phi)$, where we define
\begin{eqnarray}
\label{teq}
c_2(\phi)=\phi \{ N_c (K/\Delta E)[2\exp(-q)\\ \nonumber
+\exp(-q/2)\sqrt{2\pi q}]+
\epsilon \sqrt{2\pi q} N_c (K/\Delta E)
+\epsilon ||\partial_{\theta'} R(\theta')||\}.
\end{eqnarray}

{\it Locality Bounds---}
The next result we need is a bound on $||W_1(\phi)W_2(\phi)-W(\phi)||$.
Note that $W_1 (\phi) W_2 (\phi)=\Theta\exp\{-\int_0^{\phi}
{\rm d}{\theta}\int_0^{\infty}
{\rm d}\tau 
\exp[-(\tau\Delta E)^2/(2q)]
[\tilde w_{\theta}^+(i\tau)-
h.c.]
\}$, where $w_{\theta}=\partial_{\theta}{\cal H}_{\theta,-\theta}$ and
where the time evolution is defined using the Hamiltonian
${\cal H}_{1;\theta,}+{\cal H}_{2;-\theta}$.  This differs from the
definition of $W$ in only one respect: the use of 
${\cal H}_{1;\theta,}+{\cal H}_{2;-\theta}$ to define the time
evolution rather than ${\cal H}_{\theta,-\theta}$.

We next recall the finite group velocity result, proven in \cite{fgv,lsmh},
that there exists a function $g(t,l)$, which depends on $J$, $R$, and
the lattice structure, such that
$||[A(t),B(0)]||\leq ||A||||B|| \sum_j g(t,l_j),$
where the sum ranges over sites $j$ which
appear in operator $B$, and $l_j=d(j,i)$ is the distance from $j$ to the closest site
$i$ in the operator $A$.
It was shown that there exists some constant $c_1$ such for
$|t|\leq c_1 l$,
$g(c_1 l,l)$ is exponentially decaying in $l$ for large $l$ with
correlation length $\xi_C$.

Now, we pick $B={\cal H}_{\theta,-\theta}-{\cal H}_{1;\theta}-
{\cal H}_{2;-\theta}$, with $||B||\leq 2 N_c J$.
  The operator
$B$ can be written as a sum of ${\cal H}^i$, with each such ${\cal H}^i$ having
at most $S$ sites.  Then,
we can bound the difference between the two different
definitions of $\tilde w_{\theta}^+(i\tau)$ using the two different
definitions of time evolution.  We have
\begin{eqnarray}
(2\pi)^{-1}|\int {\rm d}t (it+\tau)^{-1} \exp[-(t\Delta E)^2/(2q)]
 \{\exp[i{\cal H}_{\theta,-\theta}t]
w_{\theta} \exp[-i{\cal H}_{\theta,-\theta}t]- \\ \nonumber
\exp[i({\cal H}_{\theta,-\theta}-B)t]
w_{\theta} \exp[-i({\cal H}_{\theta,-\theta}-B)t]\}| \\ \nonumber
\leq
(2\pi)^{-1} N_c ||w_{\theta}|| \int {\rm d}t (it+\tau)^{-1}
\exp[-(t\Delta E)^2/(2q)]
\int_0^t {\rm d}t' 2 J S g(t',L/4-2R).
\end{eqnarray}

The integral $\int_0^t {\rm d}t'
2 J S g(t',L/4-2R)$ is bounded by $g(t,L/4)$\cite{lsmh}.  Thus, we find that
the difference using the two different definitions of time evolution is
bounded by $(2\pi)^{-1} N_c ||w_{\theta}||
\int {\rm d}t (it+\tau)^{-1} \exp[-(t\Delta E)^2/(2q)] g(t,L/4)$.
To bound this integral,
for $t<c_1L/4$, we can use the locality bound, while for $t>c_1L/4$, we
can use the Gaussian in $t$.  The result gives a bound
$(2\pi)^{-1}2 N_c ||w_{\theta}|| \{g(c_1 L/4,L/4)+2[\sqrt{2\pi q}/(\Delta E c_1 
L/4)]
\exp[-(c_1 L\Delta E/4)^2/(2q)]\}$.  Integrating over
$\tau$, and using $||w_{\theta}||\leq N_c K$,
we find
$||W_1(\phi)W_2(\phi)-W(\phi)||\leq c_3(\phi)$, where we define
\begin{eqnarray}
\label{cbeq}
c_3(\phi)= \sqrt{2\pi q} (K/\Delta E)
(\phi/2\pi) 2N_c^2 \times \\ \nonumber
\{g(c_1 L/4,L/4)+2\frac{\sqrt{2\pi q}}{\Delta E c_1 
L/4}
\exp[-(c_1 L\Delta E/4)^2/(2q)]\}
\end{eqnarray}

{\it Bound on Energy---}
We compute $\sum_i\langle W_1^{-n} {\cal H}^i W_1^n - {\cal H}^i\rangle$.
Note that $W_1(-\theta)=W_1^{\dagger}(\theta)$.
First we note that for sites $i$ with $L/4\leq x_i\leq 3L/4$,
$\langle W_1^{-n} {\cal H}^i W_1^n - {\cal H}^i\rangle=0$, 
since $W_1$ commutes with $H^i$ for such sites.

We now consider the change in expectation value of the
energy for sites $i$ with $-L/4\leq x_i<L/4$.
For such sites,
$|\langle W_1^{-n} {\cal H}^i W_1^n \rangle = \langle W_2^{-n} W_1^{-n}
{\cal H}^i W_1^{n} W_2^n \rangle|$.  
Thus, we can bound
$|\langle W_1^{-n} {\cal H}^i W_1^n \rangle -\langle {\cal H}^i \rangle|
\leq  2 J |W_1^n W_2^n\Psi_0 - \Psi_0|$.
Then, we can combine 
Eqs.~(\ref{teq},\ref{cbeq}) with the fact that
$R(2\pi n)\Psi_0=\Psi_0$ to bound
$|W_1^n W_2^n\Psi_0- \Psi_0|\leq c_2+c_3$.
Summing over all $V/2$ such $i$, we find
\be
\label{bd1}
\sum_i |\langle W_1^{-n} {\cal H}^i W_1^n \rangle -\langle {\cal H}^i 
\rangle|=E_n \leq
V J e_n/2,
\ee
where
\be
\label{end}
e_n=2 [c_2(2 \pi n)+c_3(2 \pi n)].
\ee
Finally, we pick $q$ to be equal to
$c_1 (L/2) \Delta E$ to obtain the lowest possible $E_n$, giving
the exponential decay of $e_n$ with $L$ claimed in the results.

We have thus bounded the difference in energies.  The same argument would
also work to bound the difference
$\langle W_1(-\phi) {\cal H}_{\phi,0} W_1(\phi) \rangle - \langle
{\cal H}\rangle$ for general $\phi$.

{\it Expectation of Translation Operator---}
We next consider $\langle \Psi_n|T|\Psi_n \rangle$.  We define a twisted
translation operator, $T_{\theta,\theta'}=R_1(\theta_1) R_2(\theta_2)
T$, where
$R_1(\theta_1)=\prod_{j, x_j=1}\exp[i\theta_1 (Q_j-\rho)]$ and
$R_2(\theta_2)=\prod_{j, x_j=L/2+1}\exp[i\theta_2 (Q_j-\rho)]$.
Then, 
$T_{\theta_1,\theta_2}$ is a symmetry of ${\cal H}_{\theta_1,\theta_2}$ and
$T_{\theta,-\theta}=
R(\theta)TR(-\theta)$, so that $T_{\theta,-\theta}\Psi_0(\theta,-\theta)=
z_0 \Psi_0(\theta,-\theta)$.  Since $Q_j$ is quantized, we have
$T_{2\pi n,0}=\exp[-i2\pi n \rho(V/L)]T$.

In this section, we will prove a bound on the difference
$|\langle W_1(-\phi) T_{\phi,0} W_1(\phi) \rangle z_0^{-1} 
-1|$.
Taking $\phi=2\pi n$, this will give a bound on the difference
$|\langle \Psi_n | T | \Psi_n \rangle - z_0 \exp[i 2 \pi n \rho(V/L)]|$ as
desired.  We have
$\langle W_1(-\phi) T_{\phi,0} W_1(\phi) \rangle z_0^{-1} 
=\langle W_1(-\phi) R_1(\phi)
[T W_1(\phi) T^{-1}]\rangle$.
Physically, this expectation value is close to unity for the following reason:
the operator $W_1$ inserts a twist between $x=0$ and $x=1$,
while $T W_1(\phi) T^{-1}$ inserts the twist between $x=1$ and
$x=2$.  The difference is the rotation of the sites with $x=1$.  Thus,
$R_1(\phi)[T W_1(\phi) T^{-1}]$ is very close to the operator
$W_1(\phi)$.  We now make this argument precise.

Note that 
$\langle W_1(-\phi) R_1(\phi)
[T W_1(\phi) T^{-1}]\rangle=
\langle W_2(-\phi) W_1(-\phi) R_1(\phi)
[T W_1(\phi) T^{-1}] W_2(\phi)\rangle$.  
Using previous results, we know
that $|\Psi_0^{\dagger} W_2(-\phi) W_1(-\phi) -\Psi_0^{\dagger} R(-\phi)|
\leq c_2(\phi)+c_3(\phi)$.  Now, consider
$[T W_1(\phi) T^{-1}] W_2(\phi) \Psi_0$.
Let us define a new twisted
Hamiltonian, ${\cal H}'_{\theta_1,\theta_2}$, in which the twist
in boundary conditions is by angle $\theta_1$ between
$x=1$ and $x=2$ and by angle $\theta_2$ between $x=L/2$ and
$x=L/2+1$.  Specifically,
${\cal H}_{\theta_1,\theta_2}'=R_1(-\theta_1){\cal H}'_{\theta_1,\theta_2}
R_1(\theta_1)$.
Define
$W_1'(\phi)=\Theta\exp\{-\int_0^{\phi}
{\rm d}{\theta}\int_0^{\infty}
{\rm d}\tau 
\exp[-(\tau\Delta E)^2/(2q)]
[\tilde w_{1,\theta}'^+(i\tau)-
h.c.]
\}$, where $w_{1,\theta}'=\partial_{\theta}{\cal H}'_{\theta,0}$ and
where the time evolution is defined using the Hamiltonian 
${\cal H}'_{1;\theta}\equiv T {\cal H}_{1;\theta} T^{-1}$.
Crucially, $W_1'(\phi)=
T W_1(\phi) T^{-1}$.  Similarly, define $W'$ in analogy to
the definition of $W$, using 
$w'_{\theta}=\partial_{\theta}{\cal H}'_{\theta,-\theta}$.

Then, following the same steps as above, we can show that
$|W'(\phi)\Psi_0-R_1(-\phi)R(\phi)\Psi_0|\leq c_2(\phi)$ 
and
$||W_1'(\phi)W_2(\phi)-W'(\phi)||\leq c_3(\phi)$ so that
$|[T W_1(\phi) T^{-1}] W_2(\phi)\Psi_0-R_1(-\phi)R(\phi)\Psi_0|
\leq c_2(\phi)+c_3(\phi)$.
Thus,
$|\langle W_2(-\phi) W_1(-\phi) R_1(\phi)
[T W_1(\phi) T^{-1}] W_2(\phi)\rangle-1 | \leq 2[c_2(\phi)+c_3(\phi)]$, 
and thus
$|\langle \Psi_n | T | \Psi_n \rangle - z_0 \exp[i 2 \pi n \rho(V/L)]|\leq e_n$
as desired.
Note that in this derivation, the vanishing of the phase factors
$\langle \partial_\theta'[R_1(-\theta')R(\theta')] \rangle$ and
$Z_{00}=\langle \partial_{\theta'} R(\theta') \rangle$ is crucial; we chose
the factor of $\prod_{j,x_j=1}\exp[-i\theta\rho]$ in the definition
of $R_1$ to make this phase factor vanish and it is this factor that
led to the expectation value of $T$ for state $\Psi_n$.

{\it Discussion---}
The case of general filling fraction has been previously
considered in one dimension, and used to study magnetization 
plateaus\cite{1dpt}.  It was shown that gapped states could only exist
at integer filling fraction, while at filling fraction $\rho=p/q$ with
$p,q$ coprime there were found to be at least $q$ low energy states.
We have shown very similar behavior in higher dimensions for
$\rho (V/L)=p/q$: if $\Delta E$
remains non-vanishing and $\epsilon\rightarrow 0$ sufficiently
rapidly as $L\rightarrow \infty$ then
we find that the states $\Psi_n$, for $n=0,...,q-1$ provide $q$ degenerate
low energy states.  The nature of these $q$ states is not known {\it a priori}.
They may correspond to either discrete symmetry breaking or
to topological order.  If
{\it all} local operators, not just the specific ones considered in the
proof above, have matrix elements between the low-lying
states which vanish as $L\rightarrow \infty$, then we identify this
as topological order.
If some local operator
has non-vanishing matrix elements between the local states,
then there is long-range order in that operator\cite{loc}, and we identify this
as symmetry breaking.
It is a limitation of the proof that the ability to construct low energy states
depends on the width $V/L$ of the system.  We conjecture that if
$\rho=p/q$ then one can still find $q$ low lying states for arbitrary,
sufficiently large, $V/L$,
but we are not able to prove this.

The technique shown in this paper is very general, and can be used in
any case in which one can identify a quantized conserved charge, including
cases of spins transforming under higher $SU(N)$ groups.  Comparing
to \cite{vsn}, the technique here creates a
``global vison" excitation; if instead the flux is inserted along a finite
line with endpoints, this technique can create
a local ``vison" excitation
in two-dimensional spin systems
as will be discussed elsewhere.

{\it Acknowledgements---}
I thank C. Mudry for suggesting the problem of general filling factor.
This work was supported by DOE contract W-7405-ENG-36.  

\end{document}